\documentclass[11pt,a4paper]{article}
\begin{document}

\title{ Reconsideration of the Regge-Wheeler equation \footnote{E-mail of Tian:
 hua2007@126.com, tgh-2000@263.net,xinda-2002@126.com}}
\author{Guihua Tian$^{1,2}$, Shikun Wang$^{2}$ Zhao Zheng$^{3}$\\
1.School of Science, Beijing University \\
of Posts And Telecommunications. Beijing100876, China.\\2.Academy
of Mathematics and Systems Science,\\ Chinese Academy of
Sciences,(CAS) Beijing, 100080, P.R. China,\\ 3.Department of
Physics, Beijing Normal University, Beijing100875, China.}
\date{April 18, 2005}
\maketitle

\begin{abstract}
Reconsideration of the Regge-Wheeler equation is processed by
using the Painlev\'{e} coordinate and "good" timelier to define
the initial time.  We find that: the Regge-Wheeler equation could
has positive imaginary frequency.  Because the Regge-Wheeler
equation is the odd (angular) perturbation to the Schwarzschild
black hole,  the conclusion is that the Schwarzschild black hole
is unstable with respect to the rotating perturbation.

\textbf{PACC:0420-q}
\end{abstract}

Regge and Wheeler first studied the stable problem of the
Schwarzschild black hole in 1957. They divided the perturbation
into two classes: odd and even ones \cite{rw}. Later, it is found
that odd one represents really the angular perturbation to the
metric, while even one is the radial perturbation to the metric
\cite{chan}.

Regge and Wheeler studied the stable problem in the Schwarzschild
coordinates where the background metric has an apparent
singularity at the horizon $r=2m$. So, it is very difficult to
discriminate the real divergence in the perturbation at $r=2m$
from a spurious one caused by the improper choice of the
coordinates. A solution was proposed by Vishveshwara with
transforming the perturbation quantities to the Kruskal reference
frame, which is singularity-free at the horizon $r=2m$. In
Vishveshwara's work, the perturbation fields $h_{\mu
\nu}^s(t_s,r_s,\theta, \phi)$ in the Schwarzschild coordinates are
transformed into $h_{\mu \nu}^k(v,u,\theta, \phi)$ in the Kruskal
reference frame where $(t_s,r_s,\theta, \phi)$, $(v,u,\theta,
\phi)$ are the Schwarzschild, Kruskal coordinates respectively.
Vishveshwara studied the perturbation quantities actually in
Schwarzschild coordinates timelier $t_s$, that is, the initial
conditions are defined by $t_s=0$, not by $v=0$. In this very way,
the Schwarzschild black hole is proved stable by Vishveshwara.

In fact, the metric in the Schwarzschild coordinates has another
obvious drawback: the time coordinate $t$ is not proper for
analysis of the physical process at the horizon $r=2m$. The
drawback is written in many books, for which we cite the reference
\cite{mtw} of pages 824, 826 as an instance \cite{mtw}:

\textsf{\textmd{Every radial geodesic except a "set of geodesics
of measure zero" crosses the gravitational radius at $t= +\infty$
(or at $t= -\infty$ or both) , according to Figure 31.1 and the
calculations behind that figure (exercises for the student! See
Chapter 25). One therefore suspects that all physics at $r=2M$ is
consigned to $t=\pm \infty $ by reason of some unhappiness in the
choice of the Schwarzschild coordinates. A better coordinate
system, one begins to believe, will take these two "points at
infinity" and spread them out into a line in a new $(r_{new},
t_{new})$-plane; and will squeeze the "line" ($r=2M, t$ from
$-\infty $ to $+\infty $)into a single point in the $(r_{new},
t_{new})$-plane. One is more prepared to accept this tentative
conclusion and act on it because one has already seen (equation
31.8) that the region covering the $(\theta, \phi)$ 2-sphere at
$r=2M$, and extending from $t= -\infty $ to $t= +\infty $, has
zero proper volume. What timelier indication could one want that
the "line" $r=2M$, $-\infty <t< +\infty $, is actually a point? }}

From the cited paragraph, it is clear that one must use the "good"
timelier other than $t_s$ for physical process at $r=2m$.

We reinforce our viewpoint again: we must choose "good"
coordinates to study the stable problem, just as done by
Vishveshwara in reference \cite{vish}; furthermore, we must choose
"good" timelier to study it, that is, to define the initial
conditions by "good" timelier other than Schwarzschild coordinate
timelier $t_s$.

There are several well-known coordinates for the Schwarzschild
black-hole, which are regular at $r=2m$, for example, the Kruskal
coordinate system, etc. Here, we main select the Painlev\'{e}
coordinate of the Schwarzschild black hole for the study of the
stability problem. The Painlev\'{e} coordinates were discovered
independent by Painlev\'{e} in 1921\cite{pain} and Kraus, Wilczck
in 1994 \cite{Krau}, \cite{Krau1}. The Painlev\'{e} metric is
\begin{eqnarray}
ds^{2} &=& -\left(1-\frac{2m}{r}\right)dt_p^{2}
-2\sqrt{\frac{2m}{r}}drdt_p+dr^{2}+r^{2}d\Omega ^{2} \nonumber\\
&=&
-dt_p^{2}+\left(dr-\sqrt{\frac{2m}{r}}dt_p\right)^{2}+r^{2}d\Omega
^{2}.\label{painwhitemetric}
\end{eqnarray}
It is obtained from Schwarzschild metric
\begin{equation}
ds^{2}=-(1-\frac{2m}{r})dt_{s}^{2}+(1-\frac{2m}{r})^{-1}dr^{2}+r^{2}
d \Omega ^{2}\label{orimetric}
\end{equation}
by the transformation
\begin{equation}
t_{p}=t_{s}-\left[2\sqrt{2mr}+2m \ln
\frac{\sqrt{r}-\sqrt{2m}}{\sqrt{r}+\sqrt{2m}}\right].\label{paintran1}
\end{equation}

The Painlev\'{e} metric is  stationary , and   regular at the
horizon \cite{Krau}, \cite{Krau1}, furthermore, its timelier $t_p$
is time-like for $r>2m$. These good qualities make it more
suitable for study of the stability of the Schwarzschild black
hole. Of course, to carry out the perturbation analysis entirely
in the Painlev\'{e} metric would be very difficult. We use the
Vishveshwara's result in the Schwarzschild metric, and transform
them to the Painlev\'{e} coordinates and by the timelier $t_p$ to
study the stability problem. Our conclusion is that the
Schwarzschild space-time is unstable. In fact, the timelier $t_p$
is not absolutely good for the analysis of physical process at
$r=2m$, that is, it has some drawback like the timelier $t_s$ in
some way. Nevertheless, we make further transformation
\begin{equation}
dR=\sqrt{\frac r{2m}}dr-dt_{p},\label{paintran1f}
\end{equation}
then
\begin{eqnarray}
ds^{2} = -dt_p^{2}+\frac {2m}r dR^{2}+r^{2}d\Omega
^{2}.\label{painwhitemetricf}
\end{eqnarray}
In this metric (\ref{painwhitemetricf}), the timelier $t_p$ is
well-defined. Just for simplicity, we first use the timelier $t_p$
in the metric (\ref{painwhitemetric}) to study the stable problem,
and prove it is unstable. Then, we transform the perturbation
fields further to the coordinates system (\ref{painwhitemetricf}),
and use the timelier $t_p$ in the metric (\ref{painwhitemetricf})
to prove  that the unstable conclusion also holds.

Here, we briefly explain the perturbation  of the Schwarzschild
black hole.

Suppose the background metric is $g_{\mu \nu}$, while the
perturbation in it is $h_{\mu \nu}$, the contract Ricci tensors
$R_{\mu \nu}$, $R_{\mu \nu}+\delta R_{\mu \nu} $ correspond the
metrics $g_{\mu \nu}$, $g_{\mu \nu}+h_{\mu \nu} $ respectively.
The non-linear perturbation equation is
\begin{equation}
\delta R_{\mu \nu}=0,\label{rab0}
\end{equation}
while the linear part  with respect to $h_{\mu \nu}$ in the
equation (\ref{rab0}) is  called the perturbation field equation.
After the consideration of the gauge freedom, the odd perturbation
is
$$
h_{\mu \nu}=\left|
\begin{array}{cccc}
0&                0&   0&  h_{0}(r)\\
0&                0&   0&  h_{1}(r)\\
0&                0&   0&  0       \\
h_{0}(r)&  h_{1}(r)&   0&  0
\end{array}
\right|   e^{-ikt_s}\left[\sin \theta \frac{\partial}{\partial
\theta}\right]P_{l}\left(\cos \theta\right),
$$
and the even one is
$$
h_{\mu \nu}=\left|
\begin{array}{cccc}
H_{0}(1-\frac{2m}{r})&    H_{1}&   0&  0\\
H_{1}&   H_{2}(1-\frac{2m}{r})^{-1}&   0&  0\\
0&                0&   r^{2}K&  0       \\
0&  0&   0&  r^{2}K \sin^{2}\theta
\end{array}
\right|   e^{-ikt_s}P_{l}\left(\cos \theta\right).
$$

Here we mainly discuss the odd perturbation.

In the odd perturbation, the linear equations of (\ref{rab0}) are
combined into one single equation, that is, the Regge-Wheeler
equation:
\begin{equation}
\frac{d^{2}Q}{dr^{*2}}+\left[k^{2}-V\right]Q=0,\label{ReggeWheeler}
\end{equation}
where the effective potential $V$ and the tortoise coordinate are
\begin{equation}
V=\left(1-\frac{2m}{r}\right)\left[\frac{l\left(l+1\right)}{r^{2}}-\frac{6m}{r^{3}}\right],
\end{equation}
\begin{equation}
r^{*}=r+2m\ln \left(\frac{r}{2m}-1\right)\label{tortoise}
\end{equation}
respectively. The perturbation field $h_{0}(r)$ and $h_{1}(r)$ are
connected with $Q$ by
\begin{equation}
h_{0}(r)=\frac{i}{k} \frac{d}{dr^{*}}\left(rQ\right)=\frac{i}{k}
\left[\left(1-\frac{2m}{r}\right)Q+r\frac{dQ}{dr^{*}}\right]\label{h0}
\end{equation}
and
\begin{equation}
h_{1}(r)=r\left(1-\frac{2m}{r}\right)^{-1}Q.\label{h1}
\end{equation}

In reference \cite{stew}, Stewart applied the Liapounoff theorem
to define dynamical stability of a black-hole. First, according
Stewart, the normal mode of the perturbation fields $h_{\mu \nu}$
have time-dependence of $e^{-ik t}$, which are bounded at the
boundaries of the event horizon $r=2m$  and the infinity
$r\rightarrow \infty $. The range of permitted frequency is
defined as the spectrum $S$ of the Schwarzschild black-hole. Then,
for the Schwarzschild black-hole, it could be obtained by  the
Liapounoff theorem that\cite{stew}:

(1)if $\exists k \in S$ with $\Im k >0$, the Schwarzschild
black-hole is dynamically unstable,

(2)if $\Im k <0$ for $\forall k \in S$ , and the normal modes are
complete, then, the Schwarzschild black-hole is dynamically
stable,

(3)if $\Im k \leq 0$ for $\forall k \in S$  , and there is at
least one real frequency $k \in S$, the linearized approach could
not decide the stability of the Schwarzschild black-hole.

As just reinforced, it depends on  by what timelier one defines
the time-dependence  $e^{-ik t}$.  Vishveshwara proved the normal
mode of the Schwarzschild black-hole could be real by the timelier
$t_s$. Even not considering the drawback of the timelier $t_s$,
the stability of the Schwarzschild black-hole is unsolved
according to this criterion of Stewart's definition. So, Price
studied the problem carefully \cite{pric} and Wald also treated
the problem further \cite{wald}. Of course, they all use the
timelier $t_s$ in their analysis of the stable problem.

Here, we will select the timelier $t_p$ to define the initial
condition and prove that $\Re k=0,\Im k>0$ is possible for the
Schwarzschild black hole in the metric (\ref{painwhitemetric}).
 The
drawback of the timelier $t_p$ in the metric
(\ref{painwhitemetric}) make it not suitable for the analysis at
$r=2m$.  We overcome this by  transforming it to the timelier
$t_p$ in the metric (\ref{painwhitemetricf}), and prove that $\Re
k=0,\Im k>0$ is also possible. Therefore, the Schwarzschild
black-hole is unstable according to this criterion of Stewart's
definition.

Just as done in the reference \cite{vish}, we get the perturbation
field quantities in the Schwarzschild metric coordinates for
simplification, then  obtain the corresponding quantities in the
metric (\ref{painwhitemetric}) of Painlev\'{e} coordinates by
transformation and study the stable problem.

By equation (\ref{paintran1}), one gets the perturbation fields
$[h^{p}_{ij}]$ in the metric (\ref{painwhitemetric}) of
Painlev\'{e} coordinates:
\begin{equation}
h_{03}^{p}=h_{03}^{s},
\end{equation}
\begin{equation}
h_{13}^{p}=h_{13}^{s}+\sqrt{\frac{2m}{r}}\left(1-\frac{2m}{r}\right)^{-1}h_{03}^{s},
\end{equation}
where
\begin{equation}
h_{03}^{s}=h_{0}e^{-ikt_s}=\frac{i}{k}
\left[(1-\frac{2m}{r})Q+r\frac{dQ}{dr^{*}}\right]e^{-ikt_s},
\end{equation}
\begin{equation}
h_{13}^{s}=h_{1}e^{-ikt_s}=r\left(1-\frac{2m}{r}\right)^{-1}Qe^{-ikt_s}
\end{equation}
(see equations (\ref{h0}) and (\ref{h1})), therefore,
\begin{equation}
h_{03}^{p}=\frac{i}{k}\left[\left(1-\frac{2m}{r}\right)Q+r\frac{dQ}{dr^{*}}\right]e^{-ikt_s}\label{h0white}
\end{equation}
\begin{equation}
h_{13}^{p}=\left[\frac{i}{k}\sqrt{\frac{2m}{r}}Q+r\left(1-\frac{2m}{r}\right)^{-1}
\left(\frac{i}{k}\sqrt{\frac{2m}{r}}\frac{dQ}{dr^{*}}+Q\right)\right]e^{-ikt_s}\label{h1white}
\end{equation}

Now, we solve the differential equation (\ref{ReggeWheeler})
\begin{equation}
\frac{d^{2}Q}{dr^{*2}}+\left[k^{2}-V\right]Q=0
\end{equation}

As proved in reference \cite{vish}, when $k=ik_{2}$, $k_{2}>0$,
the asymptotic solutions (\ref{ReggeWheeler}) as $r^{*}\rightarrow
\infty$ are $ \tilde{A}e^{\pm k_{2}r^{*}}$, and the well-behaved
one is $\tilde{A}e^{-k_{2}r^{*}}$, that is,
\begin{equation}
Q_{\infty}=\tilde{A} e^{-k_{2}r^{*}}
\end{equation}
and from the equation (\ref{ReggeWheeler}), the solution
$\tilde{A}e^{-k_{2}r^{*}}$ cannot become $Ae^{+k_{2}r^{*}}$ as
$r^{*}\rightarrow -\infty $. Therefore the asymptotic solution to
$r^{*}\rightarrow -\infty $ is \cite{vish}
\begin{equation}
Q_{2m}=Ae^{-k_{2}r^{*}}\label{infty}.
\end{equation}

It is obvious that $Q_{2m}$ is singular at the horizon $ r=2m $,
or, $r^{*}\rightarrow -\infty $. Nevertheless, it may be caused by
the ill-behaved-ness of the Schwarzschild metric (\ref{orimetric})
at the horizon $ r=2m $ and unphysical.

Substituting equation (\ref{infty}) into (\ref{h0white}) and
(\ref{h1white}), and using the transformation equation
(\ref{paintran1}), then,  at $ r=2m$, that is, $ r^{*}\rightarrow
-\infty $, one could obtain
\begin{eqnarray}
&&h^{p}_{03}(t_{p},r) = \frac{1}{k_{2}}\left[(1-\frac{2m}{r})-k_{2}r\right]Ae^{k_{2}(t_s-r^{*})}   \\
 &=& \frac{1}{k_{2}}\left[(1-\frac{2m}{r})-k_{2}r\right]Ae^{k_{2}t_{p}}
 e^{k_{2}\left[-r+2\sqrt{2mr}-4m\ln\left(1+\sqrt{\frac
r{2m}}\right)\right]}\label{h03-2}
\end{eqnarray}
and
\begin{eqnarray}
h^{p}_{13}(t_{p},r) &=&
\frac{1}{k_{2}}\sqrt{\frac{2m}{r}}Ae^{k_{2}(t_s-r^{*})}
+r\left(1-\frac{2m}{r}\right)^{-1}\left[\frac{1}{k_{2}}\sqrt{\frac{2m}{r}}(-k_{2})+1\right]
Ae^{k_{2}(t_s-r^{*})}   \\
 &=& \left[\frac{1}{k_{2}}\sqrt{\frac{2m}{r}}+r(1+\sqrt{\frac{2m}{r}})^{-1}\right]
 Ae^{k_{2}t_{p}}
 e^{k_{2}\left[-r+2\sqrt{2mr}-4m\ln\left(1+\sqrt{\frac
r{2m}}\right)\right]}.\label{h13-2}
\end{eqnarray}
From equations (\ref{h03-2}) and (\ref{h13-2}), it is easy to see
that $h^{p}_{13}$, $h^{p}_{03}$ are regular at $r=2m$ initially(at
$t_{p}=0$), subsequently, $h^{p}_{13}(0,r)$ and $h^{p}_{03}(0,r)$
are regular in $(2m,\infty)$, or $(-\infty ,+\infty)$ by $r^{*}$.
This induces that $\Im k=k_2>0$ belongs to the spectrum $S$ of the
Schwarzschild black-hole, that is, $h^{p}_{13}(t_p,r)$ and
$h^{p}_{03}(t_p,r)$ therefore runs into infinity as
$t_p\rightarrow \infty $.

If the timelier $t_p$ were suitable for analysis of physical
process at $r=2m$, we would get the conclusion that the
Schwarzschild black hole is unstability. Because the timelier
$t_p$ in metric (\ref{painwhitemetric})has drawback as the
Schwarzschild time coordinate $t_s$, the conclusion is not sure.
So, we overcome it by transforming to the timelier $t_p$ in the
metric (\ref{painwhitemetricf}). In this case, if $\Im k=k_2>0$
still belongs to the spectrum $S$ of the Schwarzschild black-hole,
we could get the unstable conclusion definitely.

By eq.(\ref{paintran1f}), the perturbation fields are transformed
as
\begin{equation}
\hat{h}_{03}^{p}=h_{03}^{p}+\sqrt{\frac{2m}{r}}h_{13}^{p}
\end{equation}
\begin{equation}
\hat{h}_{13}^{p}=\sqrt{\frac{2m}{r}}h_{13}^{p}
\end{equation}
where we denote the perturbation fields in coordinates
(\ref{painwhitemetricf}) as $\hat{h}_{\mu \nu}^p$. From
eq.(\ref{paintran1f}), we get the variable $r$ as function of the
variables $t_p, R$ by
\begin{equation}
r(t_p,R)=\sqrt[\frac13]{\frac9{8m}}\left(t_p+R\right)^{\frac23}.\label{r-R}
\end{equation}
So, $\hat{h}_{03}^{p},\ \hat{h}_{13}^{p}$ could be written as
\begin{equation}
\hat{h}_{03}^{p}(t_p,R)=h_{03}^{p}(t_p,r(t_p,R))
+\sqrt{\frac{2m}{r(t_p,R)}}h_{13}^{p}(t_p,r(t_p,R)),\label{r-R03}
\end{equation}
and
\begin{equation}
\hat{h}_{13}^{p}(t_p,R)=\sqrt{\frac{2m}{r(t_p,R)}}h_{13}^{p}(t_p,r(t_p,R))\label{r-R13}.
\end{equation}
First, at $t_p=0$, $\hat{h}_{03}^{p}(0,R),\ \hat{h}_{13}^{p}(0,R)$
all are well-defined from $r=2m$ to infinity; this again ensure
that  the spectrum $S$ of the Schwarzschild black-hole contain
$\Im k=k_2>0$. Because the timelier $t_p$ in the metric
(\ref{painwhitemetricf})is "good", we could certainly state that
the Schwarzschild
 space-time  is unstable with respect to this kind
perturbation.  The conclusion is obvious also from the fact that
$\hat{h}_{03}^{p}(t_p,R),\ \hat{h}_{13}^{p}(t_p,R)$ all grow into
divergence as $t_p\rightarrow \infty $ by
eqs.(\ref{r-R})-(\ref{r-R13}), (\ref{h03-2})-(\ref{h13-2}) .

We restate our conclusions again: the Schwarzschild space-time is
unstable with respect to the angular perturbations. Besides, our
method could be used in study of the stability problem of
Reissner-Nordstr\"{o}m black-hole, similar results could be
obtained.

\section*{Acknowledgments}

We are supported in part by the National Science Foundation of
China under Grant No.10475013, No.10373003, No.10375087,
No.10375008, NKDRTC(2004CB318000) and the post-doctor foundation
of China.

\end{document}